\documentclass[twocolumn,preprintnumbers,amsmath,amssymb,superscriptaddress,showpacs,aps]{revtex4}

\usepackage{graphicx}	
\usepackage{dcolumn}	
\usepackage{bm}		

\usepackage{colordvi}
\usepackage{epsfig}
\usepackage{graphicx}

\newcommand{\eq}[1]{Eq.(\ref{#1})}
\newcommand{\fig}[1]{Fig.~\ref{#1}}
\newcommand{\tab}[1]{Table~\ref{#1}}
\newcommand{\olcite}[1]{Ref.~\onlinecite{#1}}
\newcommand{\rhocr}{\rho_{\rm crit}}
\newcommand{\figwidth}{0.75\columnwidth}

\begin{document}

\title{Structure and phase equilibria of the Widom-Rowlinson model}

\author{J.M.~Brader}
\affiliation{Fachbereich Physik, Universit\"at Konstanz, D-78457 Konstanz, 
Germany}

\author{R.L.C.~Vink}
\affiliation{Institut f\"ur Theoretische Physik, 
Heinrich-Heine-Universit\"at D\"usseldorf, Universit\"atsstra{\ss}e 1, 
40225 D\"usseldorf, Germany}

\date{\today}

\begin{abstract}

The Widom-Rowlinson model plays an important role in the 
statistical mechanics of second order phase transitions and yet there 
currently exists no theoretical approach capable of accurately 
predicting both the 
microscopic structure and phase equilibria. 
We address this issue using computer simulation, density functional 
theory and integral equation theory.
A detailed study of the pair correlation functions obtained 
from computer simulation motivates a closure of the Ornstein-Zernike 
equations which gives a good description of the pair structure and 
locates the critical point to an accuracy of $2$\%.

\end{abstract}


\maketitle

\section{Introduction}

The Widom-Rowlinson (WR) model \cite{widomrowlinson,widomrowlinson_book} 
is the simplest model which correctly captures the phenomenology of 
fluid-fluid demixing for systems interacting via short range forces and is 
therefore of fundamental importance in the theory of fluids. In 
particular, the model undergoes phase separation at sufficiently high 
density with a critical point which belongs to the Ising universality 
class \cite{fisher}. The model can be regarded as a non-additive 
hard-sphere mixture in which like species do not interact but unlike 
species exhibit hard-sphere repulsion with a given collision diameter 
$\sigma$. 
Although there exist related WR-type models in which only the cross 
interaction is non-zero we will reserve the term WR model for that with 
a hard sphere interaction between unlike species.

Despite the simplicity of the interactions, an accurate theory of the bulk 
structure and thermodynamics of the WR model has proved elusive. The 
lowest order mean-field (MF) theory \cite{widomrowlinson, 
widomrowlinson_book} yields a crude description of the pair correlation 
functions and predicts a phase boundary between $A$ and $B$-rich phases 
for which the location of the critical point is in considerable error when 
compared to recent simulation estimates \cite{shewyethiraj, chayes, gozdz, 
buhot, fantoni, vink:2006}. However, to go beyond the lowest order theory 
appears to be a very demanding task. The first attempts at systematic 
improvement were made by incorporating information from higher order 
virial coefficients into the theory. In \cite{melnyk} virial, activity and 
cumulant expansions were considered.  In \cite{straley} the virial series 
was studied for a WR-type model in which the $AB$ interaction is that of 
oriented hard cubes (chosen to facilitate the calculation of higher order 
virial coefficients). In neither case was reliable improvement obtained. 
Including additional virial coefficients was found to yield either results 
worse than the MF theory or no critical point, due to the appearance of 
multiple van der Waals loops in the free energy. The origin of these 
difficulties is the extremely erratic behaviour of the partial sums of the 
virial series. This issue was addressed in \cite{stillinger} in which the 
virial series of a WR type model with Gaussian $AB$ Mayer function was 
studied. Even though the virial series for this model is much better 
behaved than for the WR case, Pad\'e summation of terms up to 11th order 
in density was required to locate the critical point with reasonable 
accuracy. Convergence of the virial series in the neighbourhood of the 
critical point for the WR model appears to be much slower than for the 
Gaussian version. Although caution should be exercised when drawing 
conclusions regarding critical behaviour from low density expansions, it 
seems likely that the critical point of the WR model lies outside the 
radius of convergence of the virial series.

The correlation functions of the MF approximation in both the 
two-component and effective one-component versions of the model were 
studied in \olcite{guerrero}. The correlation functions were shown to 
display classical behaviour at the MF critical point and were proven to be 
exact in the limit of infinitely high dimensionality. A number of 
established integral equation theories have also been applied to the WR 
model \cite{shewyethiraj} although none were able to account 
satisfactorily for the structure or phase boundary. Of the theories tested 
only the Percus-Yevick (PY) equation displayed a spinodal line, in all 
other cases (hypernetted chain, Martynov-Sarkisov, Rogers-Young and 
Ballone-Pastore-Galli-Gazzillo equations) the theory no longer converges 
in a region of the phase diagram, prior to divergence of the partial 
structure factors. Although PY is the best among the standard theories, it 
still gives a poor description of the structure and the critical points 
obtained from the spinodal and virial free energy are highly inconsistent. 
In order to address these difficulties Yethiraj and Stell (YS) developed an 
integral equation specifically for the WR model in which analytic 
expressions are derived for the direct correlation functions between like 
species $c_{AA}(r)$ and $c_{BB}(r)$ by resumming a class of diagrams which 
can be evaluated analytically \cite{yethirajstell,jagannathen}. 
Unfortunately the YS equation strongly overestimates the structure and the 
pair correlation functions do not compare favourably with computer 
simulation. 

In \olcite{matthias_wrfunctional} a fundamental measures density 
functional was constructed for the WR model suitable for application to 
inhomogeneous situations. 
The theory predicts reasonable behaviour for the pair correlations 
and improves slightly on the predictions of the MF theory for the 
phase behaviour. However, the location of the critical point remains 
in error and thus does not permit investigation of the 
structure at statepoints in the vicinity of the simulation critical point.
Despite all of these efforts it can be 
concluded that the overall level of accuracy of the existing theories 
remains unsatisfactory, given the fundamental nature of the model.

In this work we seek to develop a theory for the WR model which provides  
accurate predictions for the pair structure and which is able to locate 
the critical point to an acceptable level of accuracy. 
In order to base our approximations on firm foundations we will use 
simulation results for the pair structure to guide the construction of 
our theory.
We focus in particular on the the 
total correlation function $h_{ij}(r)=g_{ij}(r)-1$, where $g_{ij}(r)$ are 
the radial distribution functions, the bridge function $b_{ij}(r)$ and 
the direct correlation function $c_{ij}(r)$.
The paper is structured as follows: We begin by 
discussing the model in Section \ref{model}. In Section~\ref{simulation}, 
we present computer simulation results for $h_{ij}(r)$, $c_{ij}(r)$ and 
$b_{ij}(r)$, focusing on high density statepoints off the critical line where 
the model is least well understood. 
These simulation results will act as motivation for our theoretical 
approaches in Section~\ref{theory}. Finally, we summarize our results and 
suggest possibilities for future work.

\begin{table}

\caption{\label{table1} Properties of the statepoints at which pair 
correlation functions were obtained using computer simulation. Shown for 
each statepoint is the density $\rho_A$ of $A$ particles, the fugacity 
$z_B$ of $B$ particles, as well as the corresponding density $\rho_B$ of 
$B$ particles, the composition $x$, and the total density $\rho$.}

\begin{ruledtabular}
\begin{tabular}{ccc|ccc}
statepoint & $\rho_A$ & $z_B$ & $\rho_B$ & $x$ & $\rho$ \\ \hline
1 & 0.1 & 0.8 & 0.5815 & 0.1467 & 0.6815 \\
2 & 0.3 & 0.8 & 0.3366 & 0.4713 & 0.6366 \\
3 & 0.4 & 0.8 & 0.2481 & 0.6172 & 0.6481 \\
4 & 0.5 & 0.8 & 0.1758 & 0.7399 & 0.6758 \\
5 & 0.6 & 0.8 & 0.1191 & 0.8344 & 0.7191 \\
6 & 0.7 & 0.8 & 0.0774 & 0.9004 & 0.7774
\end{tabular}
\end{ruledtabular}

\end{table}

\section{The WR model}
\label{model}

\begin{figure}
\epsfig{file=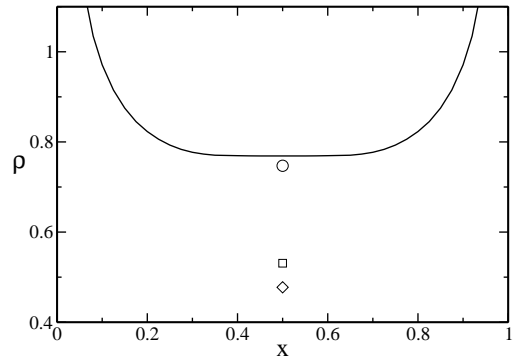,width=6cm,angle=-90}
\caption{\label{phasediagram} The phase diagram of the WR model in 
$(x,\rho)$ representation. The line is the spinodal from the new integral 
equation closure, see \eq{closure1}, of this work. The circle marks the 
location of the critical point, as obtained in the simulations of 
\olcite{vink:2006}. For comparison we also show the critical point 
predicted by the simple mean field theory of 
\cite{widomrowlinson,widomrowlinson_book} (diamond) and the density 
functional theory of \olcite{matthias_wrfunctional} (square).}
\end{figure}

The binary WR model is a symmetric mixture consisting of species $A$ and 
$B$. The interaction between like-species is ideal, $\phi_{AA}(r) = 
\phi_{BB}(r)=0$, while unlike-species interact via a hard-core potential 
of diameter $\sigma$, $\phi_{AB}(r)=\infty$ for $r<\sigma$, and zero 
otherwise. We henceforth take $\sigma$ as the unit of length. 
Above a certain 
critical density $\rho = (N_A + N_B)/V$, the WR model phase separates into 
two phases: one phase containing predominantly $A$ particles, and the 
other phase mostly $B$ particles. Here, $V$ is the volume, and $N_A$ 
($N_B$) denotes the number of $A$ ($B$) particles in the system. Due to 
the symmetry of the model, the compositions of the phases are given by $x$ 
and $1-x$, respectively, with $x=N_A/(N_A+N_B)$. At the critical point one 
has $x=1/2$. The phase diagram is thus conveniently represented in the 
$(x,\rho)$-plane, see \fig{phasediagram}. For densities $\rho > \rhocr$, 
coexisting phases of composition $x$ and $1-x$ can be identified. The 
binodal, which is symmetric about the line $x=1/2$ in the $(x,\rho)$ 
representation, terminates at the critical point. In the mean field 
approximation the binodal exhibits a parabolic curvature around the 
critical point (recall the mean-field critical exponent of the order 
parameter $\beta=1/2$). The simplest mean-field estimate of the critical 
density is $\rhocr =3/2\pi=0.4775$ 
\cite{widomrowlinson,widomrowlinson_book}, whereas the critical point of 
the PY spinodal lies at $\rhocr = 1.12$
\cite{ahn_lebowitz,PYstudy}. We emphasize 
that both of these approximations are mean field in character and exhibit 
classical critical exponents. In contrast, the best current simulation 
estimates for the critical density are $\rhocr = 0.7470(8)$ \cite{buhot} 
and $\rhocr = 0.7486 \pm 0.0002$ \cite{vink:2006}, which, as was pointed 
out in the Introduction, is not accounted for by any theoretical approach.

\begin{figure}
\begin{center}
\includegraphics[clip=,width=\figwidth]{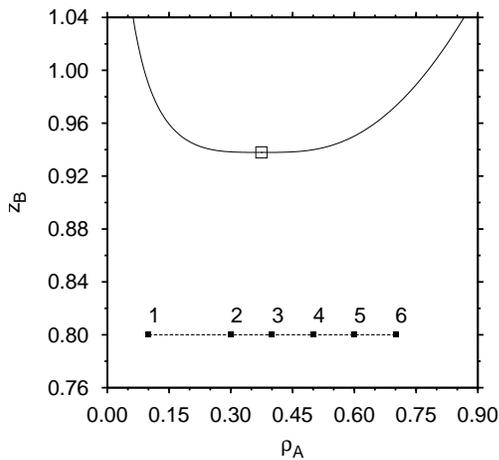}
\caption{\label{simbin} Phase diagram of the WR model in ($\rho_A,z_B$) 
representation. The solid curve shows the binodal; the open square marks 
the location of the critical point. The binodal was constructed using 
simulation data of \olcite{vink:2006} combined with finite size scaling, 
and so, on the scale of the above graph, accurately reflects the true 
thermodynamic limit form. The closed squares (labeled $1-6$ on the 
horizontal line in the one-phase region), mark the statepoints at which 
the simulations of this work were performed to obtain the pair correlation 
functions, see also \tab{table1}.}
\end{center}
\end{figure}

A possible source for the discrepancy between simulation and theory is the 
fact that the WR model belongs to the universality class of the Ising 
model. For the Ising model, $\beta \approx 0.326$ \cite{fisher.zinn:1998}, 
implying a flatter binodal. Computer simulations of the WR model indeed 
recover Ising critical behavior \cite{chayes, buhot, gozdz, vink:2006}. 
However, in order to observe the pure Ising exponent $\beta$, the 
$(x,\rho)$ representation of the binodal is not the most convenient. For 
Ising systems, there is an additional singularity in the specific heat, 
governed by the critical exponent $\alpha \approx 0.109$ 
\cite{fisher.zinn:1998}. In the $(x,\rho)$ representation, the curvature 
of the binodal is then described by the renormalized exponent $\beta^\star 
= \beta/(1-\alpha)$ \cite{fisher:1968} (in general, critical exponents 
become renormalized if the critical point is approached by varying a 
quantity which is not a field variable). In order to observe the pure 
Ising exponent, the binodal should be represented in analogy to the 
(density,temperature) phase diagram of simple fluids. For the WR model, 
this would be a grand-canonical representation, where the density 
$\rho_A=N_A/V$ of $A$ species, and the fugacity $z_B$ of $B$ species, are 
the relevant variables \cite{chayes, vink:2006, note:2006} (the choice for 
$A$ or $B$ is of course arbitrary). Shown in \fig{simbin} is the phase 
diagram in ($\rho_A,z_B$) representation. The curvature of the binodal 
around the critical point, at $\rho_A \approx 0.3743$ and $z_B \approx 
0.93791$, is now described by the pure Ising exponent $\beta$ 
\cite{vink:2006}. In contrast to the ($x,\rho$) representation, the 
symmetry of the WR model is not obvious from the binodal of \fig{simbin}. 
The symmetry, of course, still exists. In the grand-canonical ensemble, it 
corresponds to the line of equal fugacities $z_A=z_B$. Note that for 
mean-field systems, the curvature of the binodal is not affected by the 
representation, since here $\alpha=0$.

\section{Computer simulations}
\label{simulation}

Fantoni {\it et al.} \cite{fantoni} have presented computer simulation 
results for $g_{ij}(r)$ and $c_{ij}(r)$ for several different values of 
$\rho$ along the symmetry line $x=1/2$. One of the most interesting 
conclusions arising from this work is that the Percus-Yevick condition, 
$c_{AB}(r)=0$ for all $r>1$, is satisfied to very high accuracy, even 
for statepoints approaching the critical point. However, it is not at all 
clear whether this property is also maintained off the symmetry line; the 
non-trivial cancellations which apparently occur in the diagrammatic 
expansion of $c_{ij}(r>1)$ on the symmetry line may no longer hold 
when $\rho_A \ne \rho_B$. Indeed, a theoretical approach which aims to 
describe phase separation must be able to accurately describe the change 
in pair structure as a function of $x$. As far as we are aware there exists 
no detailed study of the behaviour of the pair structure for $x \ne 1/2$.

\subsection{Simulation details}

Motivated by the above considerations, we have performed simulations for 
the off-symmetry statepoints given in \tab{table1}. To simulate the 
off-symmetry statepoints $x \neq 1/2$, we use a quasi-grand canonical 
simulation ensemble, whereby the system volume $V$, the density of A 
particles $\rho_A$, and the fugacity $z_B$ of $B$ particles are fixed, 
while the number of $B$ particles fluctuates. The simulations are 
performed in cubic simulation boxes of edge $L=30$, using periodic 
boundary conditions in all $d=3$ directions. For the statepoints 
considered by us, this implies approximately 15,000 particles in each 
simulation box. To simulate efficiently, a cluster Monte Carlo move is 
used \cite{vink.horbach:2004*1, vink:cm}. We specialize to $z_B=0.8$, 
which is well below its critical value $z_{B,cr} \approx 0.93791$ 
\cite{chayes, vink:2006}, and inside the one-phase region of the phase 
diagram, see \fig{simbin}. The density of the $A$ particles is then varied 
over the range $0.1-0.7$. For each statepoint, the average concentration 
$\rho_B$ of $B$ particles is measured, as well as the radial distribution 
functions $g_{ij}(r)$. The radial distribution functions are evaluated 
using a standard method \cite{allen.tildesley:1989}, and averaged over 
approximately 5000 independent configurations. For each statepoint, this 
requires an investment of about 120 CPU hours. A total of six distinct 
statepoints is considered. For each statepoint, the average concentrations 
$\rho_A$ and $\rho_B$, as well as the total concentration $\rho$, and the 
composition $x$, are listed in \tab{table1}.

\subsection{Analysis of simulation structure}

\begin{figure}
\includegraphics[clip=,width=\figwidth,angle=270]{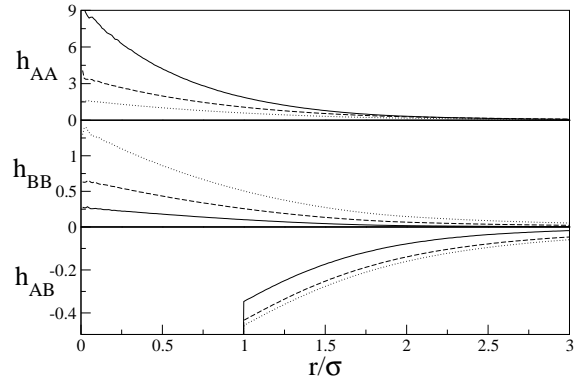}
\caption{ \label{h_structure} The correlation functions $h_{ij}(r)$ from 
simulation for statepoints $6$ (solid line), $4$ (broken line) and $2$ 
(dotted line) (see Table \ref{table1}). 
Note that $h_{AB}(r<1)=-1$.}
\end{figure}

\begin{figure}
\includegraphics[clip=,width=\figwidth,angle=270]{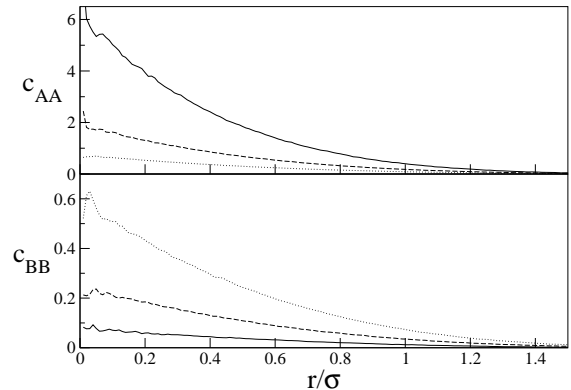}
\caption{\label{c_structure1} The correlation functions $c_{AA}(r)$ and 
$c_{BB}(r)$ from simulation for statepoints $6$ (solid line), $4$ (broken 
line) and $2$ (dotted line).}
\end{figure}

\begin{figure}
\includegraphics[clip=,width=\figwidth,angle=270]{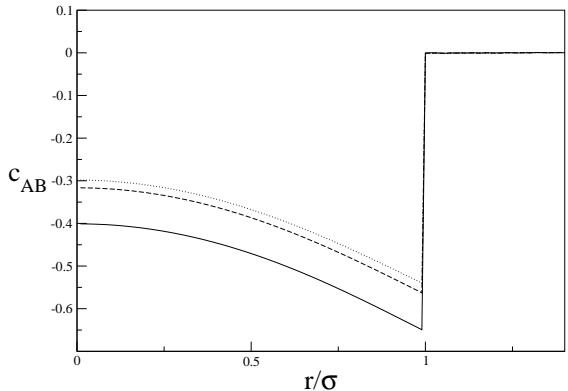}
\caption{\label{c_structure2} The correlation function $c_{AB}(r)$ from 
simulation for statepoints $6$ (solid line), $4$ (broken line) and $2$ 
(dotted line). $|c_{AB}(r>1)|<10^{-3}$ at all simulated statepoints. 
}
\end{figure}

\begin{figure}
\includegraphics[clip=,width=\figwidth,angle=270]{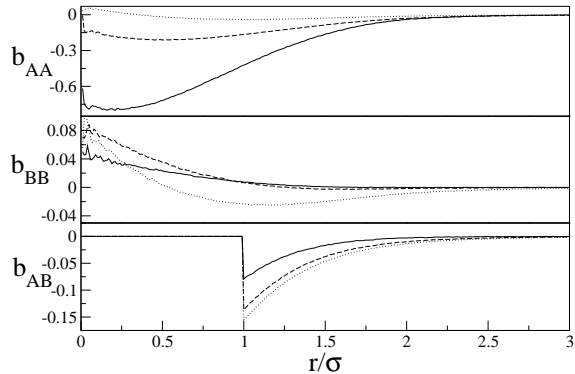}
\caption{\label{b_structure} The correlation functions $b_{ij}(r)$ from 
simulation for statepoints $6$ (solid line), $4$ (broken line) and $2$ 
(dotted line).}
\end{figure}

In \fig{h_structure} we show the total correlation functions $h_{ij}(r)$ 
for three of the considered statepoints. 
The correlations between like species are monotonic and exhibit no sign 
of any oscillatory packing behaviour. 
The increase of the $h_{ii}(r)$ as $r\rightarrow 0$ reflects the tendency 
of like species to overlap in order to maximize the free-volume and hence 
the entropy of the system. 
Equivalently, this clustering behaviour can be viewed as reflecting the attractive
(many-body) depletion potential acting between spheres of species $A$ ($B$), 
induced by the sea of non-interacting spheres of species $B$ ($A$).
For example, for small values of $x$ we expect 
$h_{AA}(r)=\exp[-\beta\phi_{dep}(r)]-1$ and 
$c_{AA}(r)=\exp[-\beta\phi_{dep}(r)]+\beta\phi_{dep}(r)-1$,
with depletion potential
\begin{equation}
\beta\phi_{dep}(r)=-\frac{4}{3}\pi\rho_B\left( 1 
- \frac{3}{4}r 
+ \frac{1}{16}r^3 \right),
\end{equation}
for $r<2$, and zero otherwise. 
The cross correlation function $h_{AB}(r)$ is negative for all values of 
$r$ and indicates that in addition to the trivial hard core exclusion there 
is also an effective repulsion between species of opposite type. This 
is a consequence of the clustering of like particles and leads to the 
appealing picture of the bare hard-sphere repulsion between species $A$ 
and $B$ being supplemented by a softly repulsive `dressed' interaction 
describing particles shrouded by a cluster of like particles.
Naturally, this effective interaction is of statistical origin and is 
therefore not to be taken too literally.
The direct correlation functions are shown in Figs.\ref{c_structure1} and 
\ref{c_structure2}. 
To obtain the $c_{ij}(r)$ from our simulated $h_{ij}(r)$ we apply 
the method described in \cite{fantoni}. 
The Fourier transform $\tilde{h}_{ij}(k)$ yields $\tilde{c}_{ij}(k)$ 
via the Ornstein-Zernike relation. 
We then construct the difference 
$\tilde{\gamma}_{ij}(k)=\tilde{h}_{ij}(k) - \tilde{c}_{ij}(k)$ (the 
Fourier transform of a continuous function in real space) 
and transform back to get $\tilde{\gamma}_{ij}(k)$. 
We thus obtain 
$c_{ij}(r)$ from the difference $h_{ij}(r) - \gamma_{ij}(r)$. 
Both $c_{AA}(r)$ and $c_{BB}(r)$ display the same 
monotonic behaviour observed for the total correlation functions. The 
$c_{ii}(r)$ are shorter range functions than the corresponding 
$h_{ii}(r)$, as expected. 
In Figure \ref{c_structure2} we show the cross correlation function 
$c_{AB}(r)$. 
The form of $c_{AB}(r)$ inside the core ($r<1$) is quite different 
from the familiar case of an additive hard sphere system for which it 
is found that $c^{hs}_{AB}(0)\le c^{hs}_{AB}(1^-)$ at all 
densities. 
Outside the core, $r>1$, we find that the value of $c_{AB}(r)$ 
does not exceed $10^{-3}$ for any of the simulated statepoints. 
A quantity which is often of interest in liquid state integral 
equation theories are the bridge functions $b_{ij}(r)$. The bridge 
functions for a binary mixture interacting via pair potentials are defined 
by the relation
\begin{equation}
h_{ij}(r) = e^{-\beta\phi_{ij}(r)+h_{ij}(r)-c_{ij}(r)+b_{ij}(r)}-1,
\label{bridgefunction}
\end{equation} 
where $\phi_{ij}(r)$ is the pair potential acting between species $i$ and 
$j$. In \fig{b_structure} we show the bridge functions obtained from our 
simulations. 
Note that data for $b_{ij}(r)$ for $r<1$ are not 
displayed as these are not required for calculations of the pair structure 
for systems with hard core interactions, see \eq{bridgefunction}. 
While the function $b_{AB}(r)$ is of a rather simple form the functions 
$b_{ii}(r)$ take both positive and negative values.

We now consider the implications of the above for constructing approximate 
theories. The most important information to come from the simulations is 
that the PY approximation $c_{AB}(r>1)=0$ is satisfied to high 
accuracy at all of the considered statepoints. This fully supports and 
extends the findings of Fantoni {\it et al.} \cite{fantoni} and implies 
that the condition $c_{AB}(r>1)=0$ should be enforced in any 
approximate theory for this model. As the exact core condition 
$h_{AB}(r<1)=-1$ is also satisfied, this effectively reduces the 
problem to that of finding accurate approximations for $c_{AA}(r)$ and 
$c_{BB}(r)$. We note that our verification of the PY condition for the 
cross correlation functions invalidates the speculation in the discussion 
of \olcite{yethirajstell}. Here, it was suggested that a more accurate 
$c_{AB}(r>1)$ could be achieved by modeling the tail with a Yukawa 
function and adjusting the free parameters to achieve thermodynamic self 
consistency. This program, while successful for the case of hard spheres, 
would apparently lead to no significant improvement over the simple PY 
approximation for the present model.


The direct correlation functions $c_{ii}(r)$ are of significantly shorter 
range than the $h_{ii}(r)$ and possess a relatively simple monotonic form. 
It may therefore be useful to model the $c_{ii}(r)$ by suitably chosen 
basis functions of finite range. By choosing finite range basis functions 
we naturally suppress the development of realistic critical exponents (the 
direct correlation functions are known to become long range at the 
critical point \cite{Fisher_criticalreview}). However, we do not 
necessarily restrict ourselves to mean field criticality in making this 
choice. The bridge functions between like species $b_{ii}(r)$ are very 
different from those of the hard sphere system \cite{hsbridge} and exhibit 
regions of both positive and negative sign. The function 
$b_{AB}(r>1)$, although superficially similar to the hard sphere 
bridge function, is found to display a quite different functional form and 
cannot be reasonably fitted using the PY bridge function for hard spheres 
at any effective density. These findings suggest that 
modified-hyper-netted-chain (MHNC) type approximations \cite{mhnc}, where 
universality of the hard-sphere family of bridge functions is assumed, 
will not prove useful in this case. Indeed, the complex damped oscillatory 
form of the functions $b_{ii}(r)$ suggests that approaches aiming to 
directly approximate the bridge functions should be avoided.

\section{Theoretical approaches}
\label{theory}

The simulation results presented in the previous section indicate that a 
successful theory for the pair structure of the WR model may be 
constructed using the conditions $h_{AB}(r<1)=-1$, 
$c_{AB}(r>1)=0$ in combination with a suitable ``ansatz'' for the 
short range functions $c_{ii}(r)$. In this section we investigate this 
possibility. Our desire to identify suitable basis functions to describe 
$c_{ii}(r)$ leads us to consider a simple virial expansion based density 
functional approximation. This yields analytic results for the pair 
structure which, although only strictly valid at low density, actually 
give a reasonable account of the structure over the entire phase diagram. 
Modification of these results to incorporate the core condition leads to a 
new integral equation closure.

\subsection{Density functional approach}
\label{DFT}

Density functional theory (DFT) is a formalism which enables the 
calculation of thermodynamic and structural properties of systems subject 
to spatial inhomogeneity \cite{bobreview}. A key result is the 
stationarity of the grand potential with respect to variations in the 
inhomogeneous density fields, $\delta\Omega/\delta\rho_i({\bf r})=0$, 
where $i$ labels the species. Given an explicit functional, this condition 
yields a set of coupled equations for the $\rho_i({\bf r})$. 
Ideally, approximations within DFT are made directly at the level of the free 
energy, which is a physically intuitive quantity. Correlation functions of 
all order can then be generated by successive functional differentiation. 
This is to be contrasted with standard integral equation approaches for 
which the closure is usually introduced at the level of the pair 
correlation functions, and which often make no direct reference to, or 
guarantee the existence of, an explicit generating functional.
In practice, the distinction between integral equation and DFT approaches 
is frequently less clear-cut and many approximate DFTs rely on 
correlation functions obtained from integral equation theories as input.

The majority of modern density functional approaches are weighted density 
approximations in which the inhomogeneous density distributions are 
smoothed by some physically motivated set of weight functions 
\cite{bobreview}. 
Using cluster expansion methods \cite{stell_review} the exact excess Helmholtz 
free energy density can be expressed as a power series in the 
inhomogeneous density fields, $\rho_i({\bf r})$. 
For the WR model, truncation of this series at $\mathcal{O}(\rho^2)$ recovers 
the original MF theory \cite{widomrowlinson,widomrowlinson_book}. 
The only diagram contributing to $\mathcal{O}(\rho^3)$ is the triangle diagram 
consisting of two root points and a single field point, all connected by Mayer 
bonds. 
As the root and field points cannot be labelled according to species without 
either an A-A or B-B Mayer bond occuring, the diagram is equal to zero. 
The first correction to 
the MF theory comes from the term $\mathcal{O}(\rho^4)$, which contains 
only one diagram.  By neglecting terms $\mathcal{O}(\rho^5)$ and higher we 
obtain the following simple approximation
\begin{eqnarray}
\hspace{-1.0cm}
\beta\mathcal{F}^{ex}[\rho_A,\rho_B]=
-\put(5,3){\circle*{6}}\put(8,3){\line(1,0){11}}
\put(22,3){\Gray{\circle*{6}}}
\hspace{1.1cm}-
\hspace{0.5cm}
\put(5,12){\Gray{\circle*{6}}}\put(8,12){\line(1,0){11}}
\put(22,12){\circle*{6}}
\put(4.5,-2){\line(0,1){11}}\put(5,-5){\circle*{6}}
\put(22.,-2){\line(0,1){11}}\put(22,-5){\Gray{\circle*{6}}}
\put(8,-5){\line(1,0){11}}
\label{expansion}
\end{eqnarray}
where black and gray field particles are associated with the density 
fields $\rho_A(\bf{r})$ and $\rho_B(\bf{r})$, respectively, and are 
connected by Mayer bonds. Note that the above diagrams are unlabeled. To 
convert to labeled diagrams requires multiplication by the appropriate 
prefactor ($1$ and $1/4$).
 
At this point we draw on the experience of previous virial expansion 
studies of the WR model \cite{melnyk,straley,stillinger,guerrero} and 
take the truncated expansion (\ref{expansion}) as the generating functional 
for our correlation functions, at least to a first level of approximation.
We argue that the above two diagrams contain the 
dominant structural elements (basis functions) for an accurate description 
of the WR model at all densities. Our reasons for this assertion are the 
following: (i) investigations of the WR virial series suggest that 
inclusion of higher order diagrams worsens the description of the 
thermodynamic properties, (ii) the pair direct correlations and radial 
distribution functions generated from the OZ route give a reasonable 
account of existing simulation results at the simulated statepoints 
(see below), (iii) 
it can be proven that the MF theory becomes exact in the limit of infinite 
dimension \cite{lie}. The key part of the proof rests on identification of 
the four field particle diagram as the numerically dominant correction 
term to the MF theory.

It should be emphasized that although \eq{expansion} provides a reasonable 
approximation for the thermodynamic functions over a portion of the phase 
diagram, the functional (\eq{expansion}) is not a good theory for the phase 
boundary and is not intended as such. Construction of a functional which 
accurately predicts both the bulk binodal and inhomogeneous structure is a 
lofty goal which we do not pursue in the present work 
(see \cite{matthias_wrfunctional} for work in this direction). 
Here we present \eq{expansion} as a means to obtain closures at the pair 
correlation level which may be subsequently modified and improved. 
The bulk free energy obtained from the uniform density limit of \eq{expansion} 
is given by
\begin{eqnarray}
\frac{\beta F^{ex}}{V} = \frac{4}{3}\pi\rho_A\rho_B 
- \frac{34816}{181440}\pi^3\rho_A^2\rho_B^2.
\label{bulk}
\end{eqnarray}
The total Helmholtz free energy, $\beta F/V = \rho\log(\rho)-\rho + \rho 
x\log(x) + \rho(1-x)\log(1-x) + \beta F^{ex}/V$, displays two van der 
Waals loops as a function of $x$ for sufficiently large values of $\rho$ 
and is thus unable to account for the demixing transition.

\begin{figure}
\epsfig{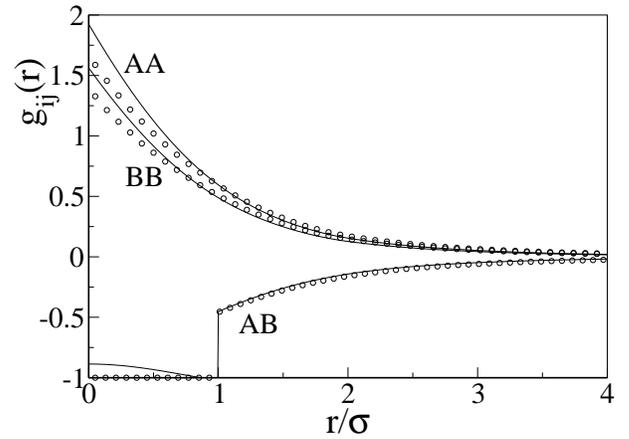}\\
\vspace{1.0cm}
\epsfig{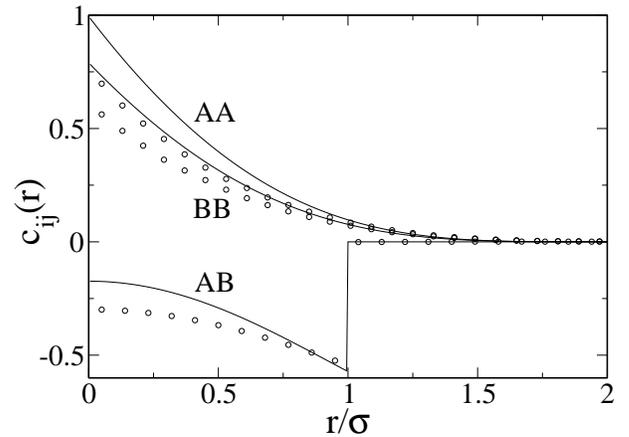}
\caption{\label{ozroutecorrelations} The total and direct correlation 
functions for statepoint $2$. Lines are the results of 
Eqs.(\ref{ozequation})-(\ref{c22}). Circles are the simulation results.}
\end{figure}

Within DFT the bulk pair correlation functions may be obtained using either 
the test-particle route (minimizing the functional in the external field due 
to a particle fixed at the origin) or the Ornstein-Zernike (OZ) route. 
Following the OZ route the inhomogeneous pair direct correlation functions 
are obtained by taking two functional derivatives of the excess free energy 
functional
\begin{equation}
c_{ij}({\bf r}_1,{\bf r}_2) = 
-\beta\frac{\delta^2 \mathcal{F}^{ex}[\{ \rho_i \}]}{\delta\rho_i({\bf r}_1)\delta\rho_j({\bf r}_2)}.
\end{equation} 
The homogeneous limit is then taken, $c_{ij}({\bf r}_1,{\bf 
r}_2)\rightarrow c_{ij}(r_{12})$, and the bulk direct correlation 
functions are substituted into the OZ relations to yield the radial 
distribution functions. For a binary fluid the OZ relations for the 
homogeneous fluid are given by
\begin{eqnarray}
\tilde{h}_{ij}(k) = \tilde{c}_{ij}(k) + \sum_l \rho_l\tilde{c}_{il}(k)\tilde{h}_{lj}(k), 
\label{ozequation} 
\end{eqnarray}
where the tilde denotes a Fourier transform. Application of this prescription to the 
functional (\ref{expansion}) generates the following simple expressions 
for the bulk direct correlation functions
\begin{eqnarray}
c_{AB}(r)&=&f(r) + \rho_A\rho_B f(r)t_2(r)\label{c12}\\
c_{AA}(r)&=&\rho_B^2 t_1^2(r)/2 \label{c11}\\
c_{BB}(r)&=&\rho_A^2 t_1^2(r)/2, \label{c22}
\end{eqnarray}
where $f(r)$ is the Mayer function, $f(r)=-1$ for $r<1$, $f(r)=0$ for 
$r>1$, and $t_1(r)$ and $t_2(r)$ are the two lowest order chain diagrams 
given by
\begin{eqnarray}
t_1(r)&=&\frac{4}{3}\pi\left( 1-\frac{3}{4}r + \frac{1}{16}r^3 \right)\,\, r\le 2 
\label{over}
\\
t_2(r)&=&-\left( \frac{\pi^2}{5040}\right) (r^3+12r^2+27r-6)(r-3)^4 \notag
\\ 
&& \hspace{3.45cm}
r\le 3, 
\end{eqnarray}
and zero otherwise. \fig{ozroutecorrelations} compares the pair 
correlation functions obtained from Eqs.(\ref{ozequation})-(\ref{c22})
with the simulation results for statepoint $2$. 
The level of agreement with the simulation 
$g_{ij}(r)$ at such a high density ($\rho \approx 0.85 \rhocr$) is 
surprising, given the fact that the correlation functions are generated 
from a truncated density expansion. 
In particular, the calculated $g_{AB}(r>1)$ lies very close to the 
simulation results. 
Although the general features of the $c_{ij}(r)$
are captured, the overall level of agreement is less satisfactory than for the 
$g_{ij}(r)$.

\begin{figure}
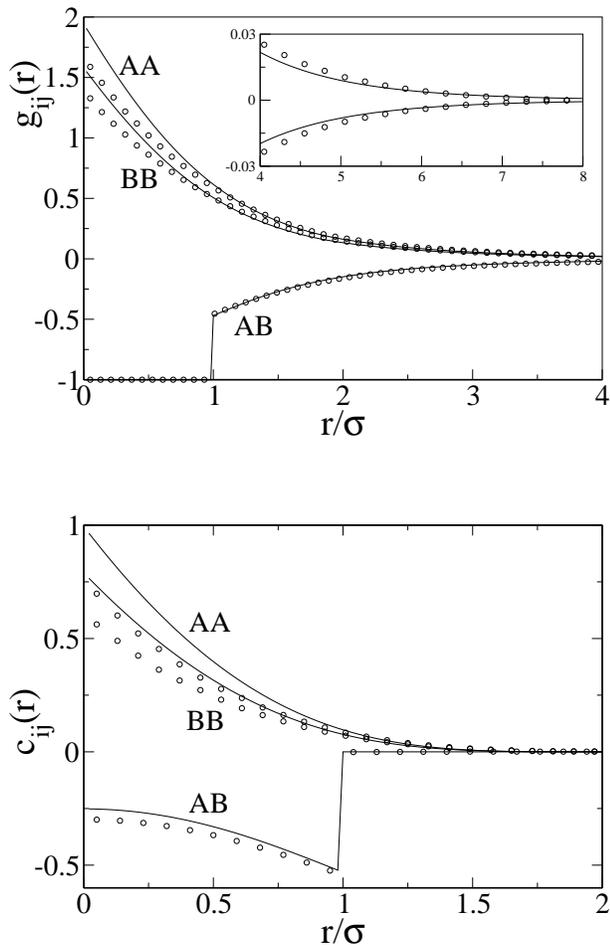

\vspace{0.cm}
\epsfig{file=figure8a.eps,width=8cm}\\
\vspace{1.0cm}
\epsfig{file=figure8b.eps,width=8cm}
\caption{\label{figclosure1} The total and direct correlation functions 
for statepoint $2$, close to the symmetry line. Lines are the results of 
closure \eq{closure1}. Circles are the simulation results.}
\end{figure}

\subsection{Imposing the core condition}

The most obvious deficiency of the present approach is the violation of 
the exact core condition, $h_{AB}(r)=-1$ for $r<1$. Violation of this 
condition is a general drawback of the OZ route in DFT studies and only in 
very special cases, e.g. the Rosenfeld functional for additive hard sphere 
mixtures \cite{rosenfeldfunctional}, is the core condition exactly 
satisfied. However, as we are primarily interested in the pair 
correlations this difficulty is easily resolved by replacing the closed 
form expression \eq{c12} for $c_{AB}(r)$ with a relation which enforces 
the core condition. Since our simulation results strongly suggest the 
approximation $c_{AB}(r>1)=0$, and given that this condition is already 
satisfied by \eq{c12}, we are led to suggest the following relations
\begin{eqnarray}
h_{AB}(r)&=&-1 \qquad r<1 \notag\\
c_{AB}(r)&=&\,\,\,\,\,0 \qquad r\ge 1 \notag\\
c_{AA}(r)&=&\rho_B^2 t_1^2(r)/2 \notag\\
c_{BB}(r)&=&\rho_A^2 t_1^2(r)/2. 
\label{closure1}
\end{eqnarray}
Combined with the OZ relation, \eq{ozequation}, this leads to a closed 
theory for the pair correlation functions. These relations correspond to a 
linearization of the expressions for $c_{AA}(r)$ and $c_{BB}(r)$ in the YS 
integral equation \cite{yethirajstell}. 


In \fig{figclosure1} we show some results obtained using closure 
(\ref{closure1}) at the same statepoint shown in 
\fig{ozroutecorrelations}. \eq{closure1} was solved using standard 
iterative numerical methods.  Imposing the core condition leads to a 
distinct improvement upon the closed form expression \eq{c12}. The 
functions $c_{AA}(r)$ and $c_{BB}(r)$ are identical to those shown in 
\fig{ozroutecorrelations}, but $c_{AB}(r)$ now lies considerably closer to 
the simulation result. The functions $g_{AA}(r)$ and $g_{BB}(r)$ remain in 
error for small separations, but the level of agreement for $r>1$ is 
improved. There are also small corrections to the function $g_{AB}(r)$ 
over the entire range. We note that by imposing the core condition we are 
effectively incorporating many more diagrams (in principle, an infinite 
number) into our description of the pair correlations. The price we pay 
for going beyond the simple virial approach of (\ref{expansion}) is that 
we must resort to fully numerical solution.

\subsection{Spinodal line and critical point}

As $\rho$ is increased for fixed $x$, the partial structure factors 
$S_{ij}(k=0)$ diverge at a well defined point. The locus of these points 
defines the spinodal line which divides the phase diagram into regions of 
mechanical stability and instability. The minimum of this curve, located 
at $x=1/2$ for the present model, identifies the critical point. It is 
well known that approximate integral equations often fail to exhibit a 
true spinodal but yield instead a no-solutions region in the phase diagram 
within which the theory simply fails to converge (see 
\olcite{brader_critical} and references therein). Indeed, the study of 
Shrew and Yethiraj \cite{shewyethiraj} performed on the symmetry line, and 
our own investigations for off-symmetry compositions, strongly suggest 
that all standard closures, with the exception of the mean-field and 
Percus-Yevick theories, fail to exhibit diverging structure factors prior 
to breakdown of the theory, and are therefore incapable of making any 
comment regarding the region in the vicinity of the critical point.

In \fig{phasediagram}, we show the spinodal resulting from the closure of 
\eq{closure1}. For comparison, we also show the critical point from the 
mean-field theory \cite{widomrowlinson,widomrowlinson_book} and that from 
the density functional theory of \cite{matthias_wrfunctional}. The 
critical point predicted by \eq{closure1} lies remarkably close to the 
simulation result. We find $\rhocr = 0.762$ which compares very favourably 
with the best current simulation estimates $\rhocr = 0.7470(8)$ 
\cite{buhot} and $\rhocr = 0.7486 \pm 0.0002$ \cite{vink:2006}. This 
represents a substantial improvement upon previous theoretical treatments. 
The numerical solution of \eq{closure1} for points of high compressibility 
(i.e. close to the spinodal) deserves some additional comment. It is a 
general difficulty of standard numerical methods based on \eq{ozequation} 
that, upon approaching the critical point, the diverging correlation 
length renders inadequate methods requiring truncation of $h_{ij}(r)$ at 
some finite range $R$. The finite size effects which result from such 
truncation give rise to considerable difficulties when attempting to 
numerically assess the critical behaviour of a given integral equation 
\cite{brader_critical}. These difficulties have been overcome for 
one-component fluids and mixtures with additive interactions using 
specialized algorithms \cite{brader_critical}. However, these methods do not 
generalize easily to non-additive mixtures, such as the WR model, and we 
have thus resorted to more traditional methods of iterative solution 
\cite{hansen}. For this reason we make no definite claims regarding the 
critical exponents of the present theory; this would require a detailed 
study using specially tailored algorithms which goes beyond the scope of the 
present work. However, the numerical methods we have employed are 
certainly capable of unambiguous determination of the spinodal line. 
This enables us to confirm that the locus of points which we have identified is 
indeed a true spinodal and not simply a region of non-convergence. 
Although we refrain from making final claims regarding the nature of the 
criticality in our equations, we do make the observation that the spinodal 
is distinctly flatter in the vicinity of the critical point, compared to 
the mean-field approaches of \cite{widomrowlinson,widomrowlinson_book, 
matthias_wrfunctional} or the PY theory. This may indicate interesting 
non-classical behaviour, and certainly warrants further investigation.

\section{Conclusions}

Using a combination of computer simulation and theoretical methods we have 
developed an integral equation for the WR model which yields good results 
for the pair structure and predicts the location of the critical point to 
an accuracy of approximately $2$\%. This represents a considerable 
improvement upon previous theories which exhibit errors in the range 
$30-50$\%. Our quasi-grand canonical computer simulations provide the 
first detailed information regarding the pair structure of the WR model 
for statepoints off the symmetry line ($x \ne 1/2$) and provide 
confirmation that the condition $c_{ij}(r>1)=0$ is satisfied to a good 
level of approximation over the entire one-phase region. The integral 
equation here developed is very simple to use and requires no more 
numerical effort than solving standard integral equations such as PY or 
HNC. Our choice of basis functions for $c_{ii}(r)$ do leave some room for 
improvement, albeit at the cost of increased numerical effort. A more 
sophisticated scheme could involve basis functions with a free parameter, 
to be determined by enforcing thermodynamic consistency between virial and 
fluctuation equations of state. Considerable success was achieved in the 
case of additive hard sphere mixtures by constructing an approximation for 
$c_{ij}(r)$ using basis functions taken from the low order diagrams in 
the virial expansion \cite{scaledfield}. 
The size of the field particle was treated as a parameter and scaled to 
interpolate between known low and high density limits. 
Whether a similar procedure is also feasible for the WR model remains an 
open question. 
Following completion of this work we were made aware of a very recent study 
in which a triplet level integral equation closure was applied to the 
WR model \cite{Malijevsky}.  
This approach is significantly more complicated than that followed in the 
present work but seems to yield very promising results worthy 
of further investigation.

\acknowledgments

This work was supported by the {\it Deutsche Forschungsgemeinschaft} under
the SFB-TR6 (project sections A6 and D3).


\begin{thebibliography}{}
\bibitem{PYstudy}
Our numerical solutions of the PY equation confirm the classical character 
of the PY critical point for the WR model, in accord with the findings 
of \cite{ahn_lebowitz}. 
We note that the PY condition $c_{ii}(r)=0$ allows  
the binary OZ equations to be reduced to a single OZ equation of standard 
form. An adaptation of the numerical method presented in \cite{brader_critical} 
can then be applied to study the critical region.
\end{thebibliography}
\end{document}